\documentclass[twocolumn]{bmcart}

\RequirePackage{natbib}
\RequirePackage[linkcolor=black,citecolor=black,urlcolor=black]{hyperref}
\usepackage[utf8]{inputenc} 

\startlocaldefs
  \usepackage{graphicx}  
  \usepackage{dsfont}    
  \providecommand{\realline}{\mathds{R}}
  \providecommand{\normaldistn}{\mathrm{N}}
  \usepackage{breakurl}  

\endlocaldefs

\begin{document}

\begin{frontmatter}
\begin{fmbox}
\dochead{Research}

\title{Hartung-Knapp-Sidik-Jonkman approach and its modification 
       for random-effects meta-analysis with few studies}

\author[
   addressref={aff1},                   
   corref={aff1},                       
   email={christian.roever@med.uni-goettingen.de}   
]{\inits{C}\fnm{Christian} \snm{R\"{o}ver}}
\author[
   addressref={aff2},
   email={guido.knapp@tu-dortmund.de}
]{\inits{G}\fnm{Guido} \snm{Knapp}}
\author[
   addressref={aff1},
   email={tim.friede@med.uni-goettingen.de}
]{\inits{T}\fnm{Tim} \snm{Friede}}

\address[id=aff1]{
  \orgname{Department of Medical Statistics, University Medical Center G\"{o}ttingen}, 
  \street{Humboldtallee~32},                     %
  \postcode{37073}                                
  \city{G\"{o}ttingen},                              
  \cny{Germany}                                    
}
\address[id=aff2]{%
  \orgname{Department of Statistics, TU Dortmund University},
  \postcode{44221}
  \city{Dortmund},
  \cny{Germany}
}

\begin{artnotes}
\end{artnotes}



\begin{abstractbox}

\begin{abstract} 
\parttitle{Background} 
  Random-effects meta-analysis is commonly performed by first deriving
  an estimate of the between-study variation, the
  \textsl{heterogeneity}, and subsequently using this as the basis for
  combining results, i.e., for estimating the \textsl{effect}, the
  figure of primary interest.  The heterogeneity variance estimate
  however is commonly associated with substantial uncertainty,
  especially in contexts where there are only few studies available,
  such as in small populations and rare diseases.

\parttitle{Methods}
  Confidence intervals and tests for the effect may be constructed via
  a simple normal approximation, or via a Student\mbox{-}$t$
  distribution, using the Hartung-Knapp-Sidik-Jonkman (HKSJ) approach,
  which additionally uses a refined estimator of variance of the
  effect estimator.
  The modified Knapp-Hartung method (mKH) applies an \textsl{ad hoc}
  correction and has been proposed to prevent counterintuitive effects
  and to yield more conservative inference.  We performed a simulation
  study to investigate the behaviour of the standard HKSJ and modified
  mKH procedures in a range of circumstances, with a focus on the
  common case of meta-analysis based on only a few studies.

\parttitle{Results}
  The standard HKSJ procedure works well when the treatment effect
  estimates to be combined are of comparable precision, but nominal
  error levels are exceeded when standard errors vary considerably
  between studies (e.g. due to variations in study size).  Application
  of the modification on the other hand yields more conservative
  results with error rates closer to the nominal level. Differences
  are most pronounced in the common case of few studies of varying
  size or precision.

\parttitle{Conclusions}
  Use of the modified mKH procedure is recommended,
  especially when only a few studies contribute to the meta-analysis
  and the involved studies' precisions (standard errors) vary.

\end{abstract}




\end{abstractbox}
\end{fmbox}

\end{frontmatter}

\section*{Background}
  Random-effects meta-analysis is most commonly performed based on an
  underlying hierarchical model including two unknowns as parameters:
  the \textsl{effect}~$\mu$, which is the figure of primary interest,
  and the \textsl{between-study variance (heterogeneity)}~$\tau^2$,
  which is a nuisance parameter. Inference then is usually done
  sequentially, by first deriving an estimate of the heterogeneity
  variance, $\hat{\tau}^2$, and then determining the effect
  estimate~$\hat{\mu}$ by conditioning on the estimate~$\hat{\tau}^2$
  \citep{HedgesOlkin,HartungKnappSinha}.  A large number of different
  estimators for the heterogeneity variance is available (see
  e.g.\ \citep{Viechtbauer2005,SidikJonkman2007,PanityakulEtAl2013,VeronikiEtAl2015}),
  and effect estimation may be done based on a simple normal
  approximation, or by utilizing a Student\mbox{-}$t$ distribution
  \citep{FollmannProschan1999} with an additionally refined estimator
  of the variance of~$\hat{\mu}$
  \citep{HartungKnapp2001a,HartungKnapp2001b,SidikJonkman2002,KnappHartung2003,InthoutIoannidisBorm2014}.
  While the normal model may be motivated by asymptotic arguments, in
  actual applications the number of estimates to be combined is
  commonly small
  \citep{TurnerEtAl2012,KontopantelisSpringateReeves2013} and hence
  the estimation uncertainty in the between-study variance~$\tau^2$ is
  substantial, so that an adjustment is appropriate and in fact
  improves operating characteristics
  \citep{FollmannProschan1999,HartungKnapp2001a,HartungKnapp2001b,SidikJonkman2002,KnappHartung2003,HigginsThompson2004,InthoutIoannidisBorm2014}.

  The problem of deriving estimates from only a small number of data
  sources is a common problem especially in fields of application
  where empirical information is sparse due to the rarity of the
  condition in question.  The rarity of a disease is often accompanied
  with a low (commercial) interest or incentive, which is why such
  diseases are also known as \textsl{orphan diseases}.  According to
  the European Commission, a disease is designated orphan status when
  the prevalence is $\leq 5$ in $10\,000$ \citep{EC1412000}.  While by
  definition any individual rare disease has a low prevalence, there
  is a large number of these, eventually affecting a substantial
  fraction of an estimated 6--8\% of the total population
  \citep{GagneEtAl2014}, and with that posing a challenge to health
  care systems worldwide.

  The European Medicines Agency acknowledges the particular obstacles
  in rare diseases research but points out that there is no
  fundamental difference between rare and more common diseases and
  hence no ``paradigm change'' when it comes to regulatory issues.
  Because of the common small-sample settings, the importance of
  sophisticated methods is emphasized, and meta-analyses of good
  quality randomised controlled clinical trials are still considered
  the highest level of evidence \citep{EMEA2006}.  The problems
  encountered in rare diseases research often call for special
  statistical methods, especially with respect to study designs
  \citep{KornEtAl2013,GagneEtAl2014,KesselheimEtAl2011}. Meta-analyses
  are particularly important in this field due to the lack of large
  trials, while these will commonly still be faced with the problem of
  small numbers of available studies.
  Between-study heterogeneity then is anticipated, since the gathered
  pieces of evidence are likely to differ with respect to study
  designs, types of control groups or treatment allocation
  \citep{KornEtAl2013,GagneEtAl2014,KesselheimEtAl2011,UnkelEtAl2015}.
  Small studies have in fact epirically been found to exhibit more
  heterogeneity than large trials \citep{InthoutEtAl2015}.
  Consequently, the use of methods suitable for few studies and
  marginally significant findings is of crucial importance here.

  With an estimated incidence of 2--20 cases per $100\,000$
  population, juvenile idiopathic arthritis (JIA) is an example of a
  rare disease \citep{PrakkenAlbaniMartini2011}. In the following, we
  will use a meta-analysis in JIA \citep{HinksEtAl2010} as a case
  study to illustrate the different methods discussed below.

  In the following sections, we will first describe the methods used,
  then show the results of a simulation study, and demonstrate the
  different types of analyses in an example data set, before closing
  with conclusions and recommendations.

\section*{Methods}
  \subsection*{Random-effects meta-analysis}
  Meta-analysis is very commonly performed via a
  \textsl{random-effects} approach, utilizing the
  \textsl{normal-normal hierarchical model}. Here the data are given
  in terms of a number~$k$ of estimates~$y_i\in\realline$ that are
  associated with some uncertainty given through standard errors~$s_i
  > 0$ that are taken to be known without uncertainty.  The estimates
  are assumed to measure trial-specific
  parameters~$\theta_i\in\realline$:
  \begin{equation}
    y_i \sim \normaldistn(\theta_i, s_i^2) \qquad \mbox{for } i = 1,\ldots,k.
  \end{equation}
  The parameters $\theta_i$ vary from trial to trial around a global
  mean~$\mu \in \realline$ due to some \textsl{heterogeneity variance}
  between trials that constitutes an additive variance component to
  the model,
  \begin{equation}
    \theta_i \sim \normaldistn(\mu, \tau^2) \qquad \mbox{for } i = 1,\ldots,k,
  \end{equation}
  where $\tau^2 \geq 0$.
  The model may then be simplified by integrating out the
  parameters~$\theta_i$, leading to the marginal expression
  \begin{equation}
    y_i \sim \normaldistn(\mu,\, s_i^2 + \tau^2) \qquad \mbox{for } i = 1,\ldots,k.
  \end{equation}
  Among the two unknowns in the model, the overall mean~$\mu$, the
  \textsl{effect}, usually is the figure of primary interest, while
  the heterogeneity variance~$\tau^2$ constitutes a nuisance
  parameter.  When $\tau^2=0$, the model simplifies to the so-called
  \textsl{fixed-effect} model \citep{HedgesOlkin,HartungKnappSinha}.

  The ``relative amount of heterogeneity'' in a meta-analysis may be
  expressed in terms of the measure~$I^2$, which is defined as
  \begin{equation}
    I^2 = \frac{\tau^2}{\tau^2+\tilde{s}^2}
  \end{equation}
  where $\tilde{s}$ is some kind of ``average'' standard error among
  the study-specific~$s_i$ \citep{HigginsThompson2002}.  In the
  following simulation studies, we will determine $\tilde{s}^2$
  as the arithmetic mean of squared standard errors.

\subsection*{Parameter estimation}
  If the value of the heterogeneity variance parameter~$\tau^2$ were
  known, the (conditional) maximum-likelihood effect estimate would
  result as the weighted average
  \begin{equation}
    \hat{\mu} = \frac{\sum_i w_i \,y_i}{\sum_i w_i}
  \end{equation}
  with ``inverse variance weights'' defined as
  \begin{equation}
    w_i = \frac{1}{s_i^2 + \tau^2} \qquad \mbox{for } i = 1,\ldots,k.
  \end{equation}
  A common approach to inference within the random-effects model is to
  first estimate the heterogeneity variance~$\tau^2$, and subsequently
  estimate the effect~$\mu$ \textsl{conditional on the heterogeneity
  estimate}.
  Note that the $w_i$ are effectively treated as ``known'' while in
  fact both the $s_i^2$ as well as $\tau^2$ are only measured with
  some uncertainty (that depends on the size/precision of the $i$th
  individual study and the number of studies~$k$).
  There is a wide range of different heterogeneity estimators
  available (see
  e.g.~\citep{Viechtbauer2005,SidikJonkman2007,PanityakulEtAl2013,VeronikiEtAl2015}
  for more details).  In the following we will concentrate on some of
  the most common ones, the DerSimonian-Laird (DL) estimator, a moment
  estimator \citep{DerSimonianLaird1986} with acknowledged
  shortcomings \citep{BoehningEtAl2002}, the restricted maximum
  likelihood (REML) estimator \citep{Viechtbauer2005,Raudenbush2009},
  and the Paule-Mandel (PM) estimator, an essentially heuristic
  approach \citep{PauleMandel1982,RukhinBiggerstaffVangel2000}.
  Software to compute the different estimates is provided e.g.\ in the
  \texttt{metafor} and \texttt{meta} \textsf{R}~packages
  \citep{Viechtbauer2010,meta}.

\subsection*{Confidence intervals and tests}
  \subsubsection*{Normal approximation}
    Confidence intervals and, equivalently, tests for the effect~$\mu$
    are commonly constructed using a normal approximation for the
    estimate~$\hat{\mu}$.  The standard error of~$\hat{\mu}$,
    conditional on a fixed heterogeneity variance value~$\tau^2$, is
    given by
    \begin{equation}
      \hat{\sigma}_\mu = \sqrt{\frac{1}{\sum_i w_i}}.
    \end{equation}
    A confidence interval for the effect~$\mu$ then results via a
    normal approximation as
    \begin{equation}\label{eqn:normalCI}
      \hat{\mu} \; \pm \; \hat{\sigma}_\mu \; z_{(1-\alpha/2)}
    \end{equation}
    where $z_{(1-\alpha/2)}$ is the $(1\!-\!\alpha/2)$-quantile of the
    standard normal distribution, and $(1\!-\!\alpha)$ is the nominal
    coverage probability \citep{HedgesOlkin,HartungKnappSinha}.  The
    normal approximation based on a heterogeneity variance
    estimate~$\hat{\tau}^2$ usually works well for many studies
    (large~$k$) and small standard errors (small~$s_i$), or negligible
    heterogeneity (small variance~$\tau^2$), but tends to be
    anticonservative otherwise
    \citep{HartungKnapp2001a,HartungKnapp2001b,SimulationPaperDummyRef}.

  \subsubsection*{The Hartung-Knapp-Sidik-Jonkman (HKSJ) method}
    Hartung and Knapp \citep{HartungKnapp2001a,HartungKnapp2001b} and
    Sidik and Jonkman \citep{SidikJonkman2002} independently
    introduced an adjusted confidence interval.  In order to determine
    the adjusted interval, first the quadratic form
    \begin{equation} \label{eqn:q}
      q = \frac{1}{k-1} \sum_i w_i(y_i-\hat{\mu})^2
    \end{equation}
    is computed \citep{HartungKnapp2001a,HartungKnapp2001b,SidikJonkman2002}.
    The adjusted confidence interval then results as
    \begin{equation}\label{eqn:knhaCI}
      \hat{\mu} \; \pm \; \sqrt{q} \; \hat{\sigma}_\mu \; t_{(k-1); (1-\alpha/2)}
    \end{equation}
    where $t_{(k-1); (1-\alpha/2)}$ is the $(1\!-\!\alpha/2)$-quantile
    of the Student\mbox{-}$t$ distribution with $(k\!-\!1)$ degrees of
    freedom.
    Note that $q \hat{\sigma}^2\mu$ is derived from a non-negative and
    unbiased estimator of $\frac{1}{\sum_i w_i}$ \citep{Hartung1999}.
    Confidence intervals based on the normal approximation may easily
    be converted to HKSJ-adjusted ones
    \citep{InthoutIoannidisBorm2014}.  The method has also been
    generalized to the cases of multivariate meta-analysis and
    meta-regeression \citep{JacksonRiley2014}.

  \subsubsection*{The modified Knapp-Hartung (mKH) method}
    The HKSJ confidence interval (\ref{eqn:knhaCI}) tends to be wider
    than the one based on the normal approximation
    (\ref{eqn:normalCI}), since the Student\mbox{-}$t$ quantile is
    larger than the corresponding normal quantile, while $q$~will tend
    to be somewhere around unity. However, $q$~may in fact also turn
    out arbitrarily small, and if
    $\sqrt{q}<\frac{z_{(1-\alpha/2)}}{t_{(k-1);(1-\alpha/2)}}$, then
    the modified interval will be shorter than the normal one, which
    may be considered counter-intuitive.  A simple \textsl{ad~hoc}
    modification to the procedure results from defining
    \begin{equation} \label{eqn:qstar}
      q^\star = \max\{1, q \}
    \end{equation}
    \citep{KnappHartung2003} and using~$q^\star$ instead of~$q$ to
    construct confidence intervals and tests.  This will ensure a more
    conservative procedure.  The modification was originally proposed
    in the meta-regression context, but the simple meta-analysis here
    constitutes the special case of an ``intercept-only'' regression.

    Note that the PM heterogeneity variance estimator is effectively
    defined by choosing $\hat{\tau}^2$ such that $q$
    (equation~(\ref{eqn:q})) is~$=\!1$ (or less, if no solution
    exists) \citep{PauleMandel1982,RukhinBiggerstaffVangel2000}, so
    that for the PM estimator the corresponding $q^\star$~value always
    equals $q^\star\!=\!1$.

\subsection*{Simulations}
  Since a $(1\!-\!\alpha)$~confidence interval is supposed to cover
  the true parameter value with probability $(1\!-\!\alpha)$, the
  calibration of such intervals may be checked by repeatedly
  generating random data based on known parameter values and then
  determining the empirical frequency with which true values are
  actually covered \citep{Dawid1982}. We performed such a
  \textsl{Monte Carlo} simulation comparing the HKSJ and mKH
  approaches using the setup that was introduced by IntHout et
  al.\ \citep{InthoutIoannidisBorm2014}.
  Data were simulated on a continuous scale, according to the
  random-effects model described above, with study-specific standard
  errors~$s_i$ set to reflect certain scenarios with respect to the
  relative size of studies and their variation due to estimation
  uncertainty.
  Many meta-analyses are based on (discrete) count data, where the
  random-effects model assumptions only hold to an approximation that
  works well unless event probabilities are very low. Alternative
  methods have been proposed to deal with low event probabilities
  \citep{BradburnEtAl2007,Kuss2015}, but low-event-rate effects were
  not considered in the present investigation.
  \onecolumn
  \begin{figure}[h!]
    \includegraphics[width=0.95\linewidth]{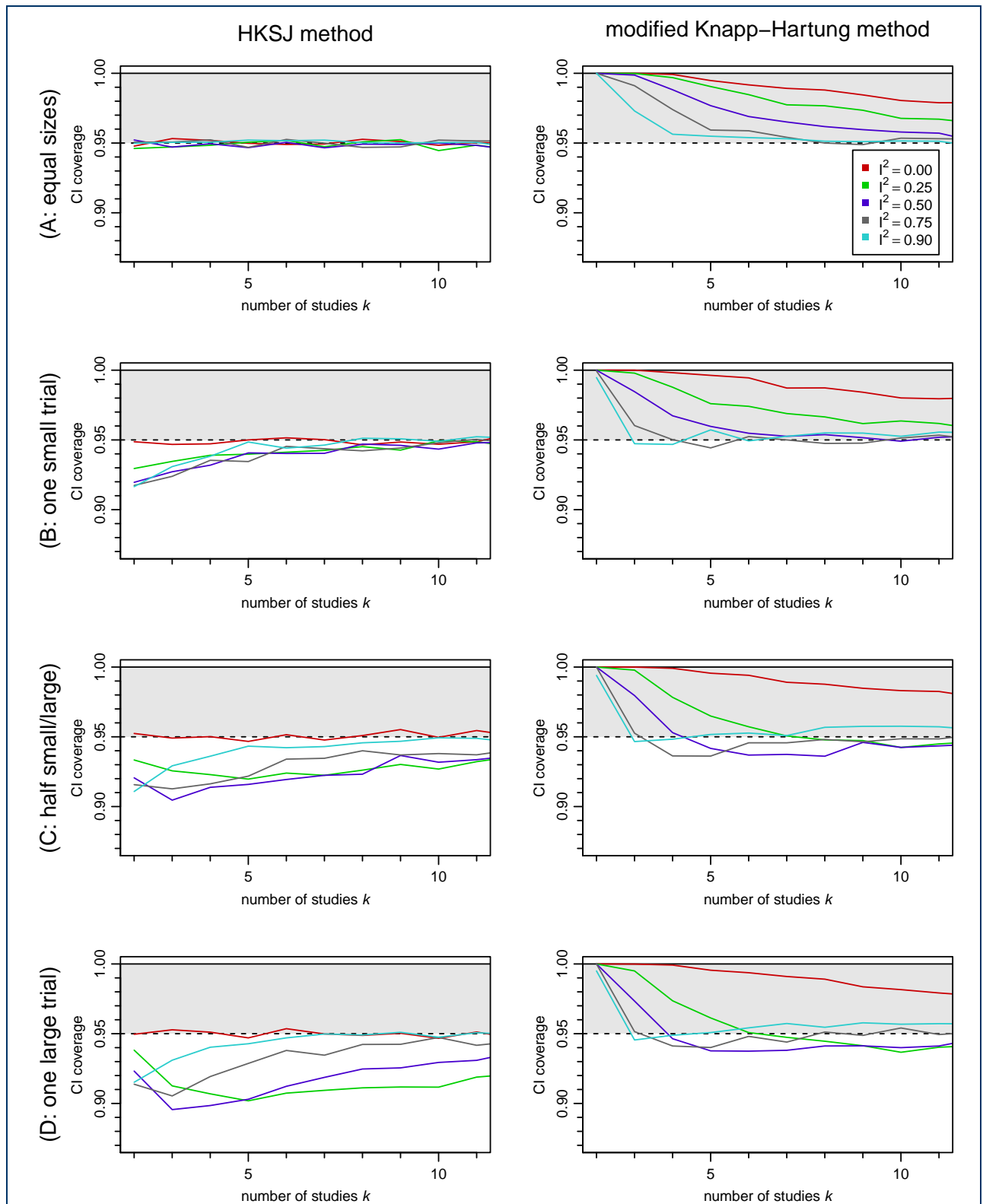}
    \caption{\label{fig:coverageDL}\csentence{Coverage probabilities
        of HKSJ and mKH 95\% confidence intervals.}  Probabilities are
      shown in dependence of the number~$k$ of studies and the amount
      of heterogeneity~$I^2$. The four different scenarios A--D
      correspond to different amounts of imbalance between the
      study-specific standard errors~$s_i$.  The DerSimonian-Laird
      (DL) method was used for estimation of the
      heterogeneity variance~$\tau^2$.}
  \end{figure}
  \twocolumn
  \noindent These simulations considered four different scenarios, namely
  meta-analyses (A) with trials of equal size, (B) with equally sized
  trials but including one small trial, (C) with 50\% large and small
  trials, and (D) equally sized trials and one large trial. The sizes
  of ``small'' and ``large'' trials (and hence, squared standard
  errors~$s_i^2$) here differ by a factor of ten, so that the
  associated standard errors differ by roughly a factor of~3; for more
  details see~\citep[Appendix~2]{InthoutIoannidisBorm2014}.  Numbers
  of studies~$k$ considered here are in the range of 2--11, and the
  true levels of heterogeneity were $I^2\in\{0.00, 0.25, 0.50, 0.75,
  0.90\}$.  At each combination of parameters $10\,000$~meta-analyses
  were simulated.
  All simulations were performed using~\textsf{R} \citep{RManual2014}.

\section*{Results}
\subsection*{Simulations}
  Figure~\ref{fig:coverageDL} shows the estimated coverage
  probabilities of the different confidence intervals based on the DL
  heterogeneity variance estimate.  The corresponding figure for REML
  and PM look essentially the same, which is in line with the findings
  in \citep{SanchezMecaMarinMartinez2008,SimulationPaperDummyRef}.
  The HKSJ method works very well when the analyzed studies are of
  equal size (i.e., have equal standard errors), as can also be shown
  analytically \citep{SidikJonkman2004},
  but coverage decreases in more imbalanced settings, especially for
  small numbers of studies.  For the case of no heterogeneity
  ($I^2=0$) the HKSJ method also works fine, but if $\tau^2$ was known
  a~priori, in this case the fixed-effect model should work as well.
  The mKH procedure on the other hand is rather conservative for
  small~$k$, but does not tend to inflate the type-I error
  substantially regardless of the underlying study sizes or true
  heterogeneity.  For both methods, the dependence on the amount of
  heterogeneity is mostly a matter of whether $\tau^2$ is~$=0$
  or~$>0$.

\begin{figure}[t]
    \includegraphics[width=0.95\linewidth]{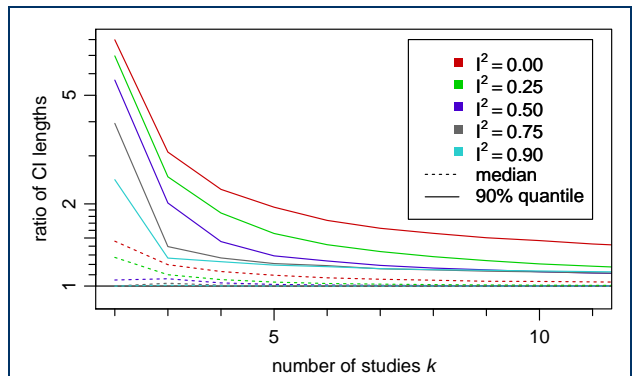}
    \caption{\label{fig:lengthMP}\csentence{Ratios of lengths of
        HKSJ and mKH confidence intervals.}
      (Same as $\frac{\sqrt{q^\star}}{\sqrt{q}}$).
      Lengths are shown in dependence of the number~$k$ of studies and
      the amount of heterogeneity~$I^2$. Numbers are averaged over all 4 scenarios. 
      The DL~method was used for $\tau^2$~estimation.}
  \end{figure}

  Application of the modification obviously tends to widen the
  resulting confidence intervals. The ratio of interval lengths (which
  is equal to $\frac{\sqrt{q^\star}}{\sqrt{q}}$) is shown in
  Figure~\ref{fig:lengthMP}. Most notably, the effect is largest for
  small heterogeneity and for few studies.

  The modification eventually only makes a difference in those cases
  where $q$~turns out smaller than one.  The fraction of intervals
  affected by the modification ranges between 31\% and 82\% for the DL
  heterogeneity variance estimator and for the scenarios investigated
  here, with an overall average of 61\%.  For the other two estimators
  the fractions are 29--82\% with a mean of~62\% (REML), and 31--91\%
  with a mean of~74\% (PM).  Again, the differences between the
  different estimators are rather small.  With respect to the
  underlying simulation scenario, the probability decreases with
  increasing heterogeneity, since larger heterogeneity also leads to
  larger values of~$q$.

\subsection*{Application to JIA example data}
\begin{figure}[t]
    \includegraphics[width=0.95\linewidth]{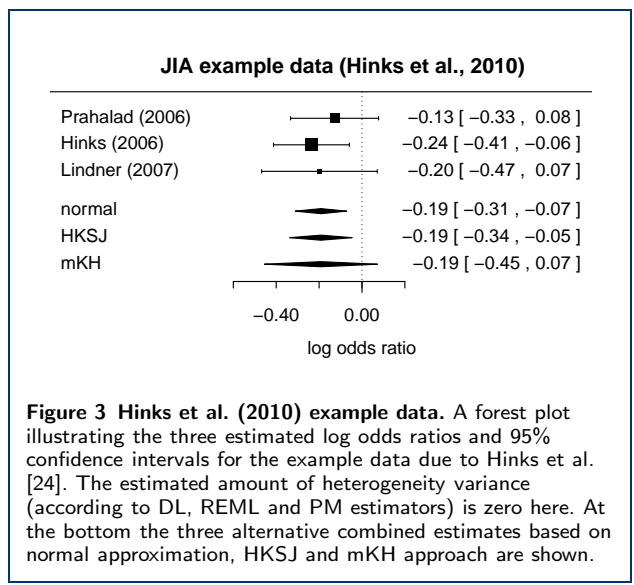}
    \caption{\label{fig:HinksForest}\csentence{Hinks et al.\ (2010)
        example data.}  A forest plot illustrating the three estimated
      log odds ratios and 95\% confidence intervals for the example
      data due to Hinks et al.\ \citep{HinksEtAl2010}.  The estimated
      amount of heterogeneity variance (according to DL, REML and PM
      estimators) is zero here. At the bottom the three alternative
      combined estimates based on normal approximation, HKSJ and mKH
      approach are shown.}
  \end{figure}

  Hinks et al.\ \citep{HinksEtAl2010} studied the occurrence of a
  particular genetic variant, CCR5, in juvenile idiopathic arthritis
  (JIA) patients in comparison with the general population. Their
  investigation included a meta-analysis of a small number ($k=3$) of
  available controlled studies looking into the association of JIA
  with this particular biomarker. The analysis was based on
  logarithmic odds ratios; the three estimates along with their
  standard errors are shown in a forest plot in
  Figure~\ref{fig:HinksForest}.
  Here the largest standard error is 50\%~larger than the smallest
  one.  For these data, all three (DL, REML and PM) heterogeneity
  variance estimates turn out as $\hat{\tau}^2\!=\!0$. A zero
  heterogeneity variance estimate is not uncommon, even when the
  actual heterogeneity is in fact substantial
  \citep{KontopantelisSpringateReeves2013}.  The associated $q$~value
  is also small with~$q=0.31$ ($\sqrt{q}=0.55$).  The resulting three
  confidence intervals based on normal approximation, HKSJ adjustment
  and mKH method all differ in their lengths. The mKH interval is
  longest, and includes the zero log odds ratio (indicating no
  association between genetic marker and disease), while the other two
  intervals do not include zero.  So the choice of procedure directly
  affects conclusions in this example.

  In this example, the three standard errors are rather similar (the
  largest is 50\% larger than the smallest), but while the
  heterogeneity variance estimates~$\hat{\tau}^2$ are all at zero, the
  95\% Q-profile confidence interval \citep{Viechtbauer2007} ranges up
  to $\tau^2=0.33^2$, corresponding to $I^2=0.90$.  Within the context
  of our simulations (Fig.~\ref{fig:coverageDL}), we cannot tell from
  the data which of the heterogeneity ($I^2$) scenarios we are in fact
  in, and as the standard errors are not exactly the same, it remains
  a matter of balancing the potential consequences whether one would
  rather risk losing on the side of (type~I) error probability or
  power.

\section*{Discussion}
  The HKSJ procedure ensures the coverage probability only when the
  included studies' standard errors~$s_i$ are similar; for unbalanced
  settings, the actual error probability tends to exceed the targeted
  one.  With the standard definition of the correction factor~$q$ the
  results may sometimes be counterintuitive, since the corresponding
  CIs may turn out shorter than using the simple normal approximation;
  in fact they may get arbitrarily short.  In case of no heterogeneity
  ($\tau=0$) the HKSJ method also works well, however practically this
  is of limited relevance, as one can rarely tell (or convincingly
  argue) whether this condition holds.

  The \textsl{ad hoc} modification of the mKH method aims at fixing
  these shortcomings and results in type-I error probabilities that
  are not grossly in excess of the pre-specified ones.
  Especially when the standard errors~$s_i$ are of dissimilar
  magnitude, the mKH method can therefore be recommended.
  For few studies (small~$k$), the modified procedure however tends to
  be very conservative, with very small error probabilities especially
  in the extreme case of meta-analysis of only $k\!=\!2$ studies.
  In this extreme case the choice of methods may therefore be
  considered a matter of a power vs.\ type\mbox{-}I error probability
  tradeoff.

  While meta-analyses of few studies are a particular problem in
  indications where there is only little evidence available (such as
  rare diseases), such circumstances are not as uncommon as one might
  expect. Turner et al.\ \citep{TurnerEtAl2012} and Kontopantelis et
  al.\ \citep{KontopantelisSpringateReeves2013} investigated the
  analyses archived in the Cochrane Database and actually found a
  \textsl{majority} of them to be based on as few as $k\!=\!2$ or
  $k\!=\!3$ studies; so these constitute highly relevant cases for
  which the proper control of error rates is crucial.
 
  The properties of either unmodified or modified method for the
  extreme case of $k\!=\!2$ may be considered unsatisfactory, as it
  seems one has the choice of either falling short of or exceeding the
  targeted error probability; the problem has in fact been regarded as
  effectively unsolved \citep{GonnnermannEtAl2015}.  The poor
  behaviour may be explained by the fact that performing a
  random-effects meta-analysis effectively means the estimation of
  first- and second-order statistics, and it is not overly surprising
  to find that this is a hard task when the data consist of as few as two
  samples that are only measured with uncertainty.  Bearing this in
  mind, the use of Bayesian methods \citep{SuttonAbrams2001} and the
  consideration of external evidence on the likely magnitude of the
  heterogeneity \citep{TurnerEtAl2015} may be the way forward.


\begin{backmatter}

\section*{List of abbreviations}
  \begin{tabular}{ll}
    DL    &  DerSimonian-Laird \\
    HKSJ  &  Hartung-Knapp-Sidik-Jonkman\\
    JIA   &  juvenile idiopathic arthritis\\
    mKH   &  modified Knapp-Hartung\\
    PM    &  Paule-Mandel\\
    REML  &  restricted maximum likelihood
  \end{tabular}

\section*{Competing interests}
  The authors declare that they have no competing interests.

\section*{Authors' contributions}
  TF conceived the concept of this study.
  CR carried out the simulations and drafted the manuscricpt. 
  GK critically reviewed and made substantial contributions to the manuscript.
  All authors commented on and approved the final manuscript.

\section*{Acknowledgements}
  This work has received funding from the European Union's Seventh
  Framework Programme for research, technological development and
  demonstration under grant agreement number FP~HEALTH~2013-602144
  through the InSPiRe project \citep{inspireWebsite}.

\bibliographystyle{bmc-mathphys} 
\bibliography{/home/christian/literature/literature}      




\end{backmatter}
\end{document}